\documentclass[sigconf]{acmart}
\acmConference[EASE 2026]{The 30th International Conference on Evaluation and Assessment in Software Engineering}{9–12 June, 2026}{Glasgow, Scotland, United Kingdom}
\AtBeginDocument{%
  }

\acmConference[EASE 2026]{The 30th International Conference on Evaluation and Assessment in Software Engineering}{9–12 June, 2026}{Glasgow, Scotland, United Kingdom}

\begin{document}

\title{Beyond Code, We Are People: A Systematic Mapping of 25 Years \\of Literature on Soft Skills in Agile Development Teams} 


\author{Israely Lima}
\orcid{0009-0004-6128-6361}
\affiliation{%
  \institution{Federal University of Ceara}
  \city{Quixadá}
  \country{Brazil}}
\email{israelylima@alu.ufc.br}

\author{Lucas Moura Lourenço}
\affiliation{%
  \institution{Federal University of Ceara}
  \city{Quixadá}
  \country{Brazil}}
\email{lucasmlourenco.es@gmail.com}

\author{Márcio Ribeiro}
\affiliation{%
  \institution{Federal University of Alagoas}
  \city{Alagoas}
  \country{Brazil}}
\email{marcio@ic.ufal.br}

\author{Ivan Machado}
\affiliation{%
  \institution{Federal University of Bahia}
  \city{Bahia}
  \country{Brazil}}
\email{ivan.machado@ufba.br}

\author{Carla Ilane Bezerra}
\affiliation{%
  \institution{Federal University of Ceara}
  \city{Quixadá}
  \country{Brazil}}
\email{carlailane@gmail.com}

\renewcommand{\shortauthors}{Lima et al.}

\begin{abstract}
  Software development is a sociotechnical and human-centered endeavor in which human factors directly influence quality, productivity, and innovation capacity. In this context, career development in computing goes beyond technical mastery, requiring competencies that enable professionals to deal with continuous change and collaborative demands. Among these, non-technical skills (soft skills) stand out, encompassing social, emotional, and communicational dimensions essential to team effectiveness and the success of software projects. Despite their recognized importance, there is still a need for a systematic mapping of the most relevant soft skills over the past 25 years, a period marked by the adoption of agile approaches in industry. This gap limits the integration of human and technical aspects in software development. This study presents a systematic mapping of the literature, analyzing 97 studies published between January 2000 and May 2025 across major scientific databases. The results identify recurring competencies such as communication, adaptability, teamwork, and leadership, as well as their association with different roles in agile contexts. The main agile approaches adopted, particularly Scrum, are also identified, along with key gaps in the literature, such as the lack of studies on role-specific soft skills. The findings can support researchers, educators, and practitioners in designing curricula, training strategies, and organizational practices aligned with human factors, reinforcing the importance of integrating social and technical dimensions in the development of collaborative and innovative professionals.

\end{abstract}



\received{25 january 2026}
\received[revised]{ }
\received[accepted]{ }

\maketitle

\section{Introduction} \label{sec:introduction}

Software development is understood as a sociotechnical and human-centered endeavor \cite{mnkandla_balancing_nodate, lenberg2015, teixeira_integrating_2016, miranda2021}. In this context, the human factor plays a decisive role in development success, directly influencing quality, productivity, and innovation capacity \cite{matturro_soft_2015, souza2024, faquin2016}. As highlighted by Wazlawick et al. \cite{wazlawick2024}, employers and organizations increasingly value not only technical expertise but also a set of skills that enables professionals to perform effectively, reducing errors and delays in the delivery of products and services. Soft skills encompass interpersonal, social, and emotional dimensions \cite{ahmed2015soft}, such as communication, teamwork, leadership, creativity, critical thinking, and adaptability \cite{otemaier2025projeto}. From this perspective, software projects can achieve better outcomes when individuals with specific skills are allocated to different phases of the project \cite{john2005, ahmed2012}. This view is further reinforced by agile processes, which emphasize continuous adaptation and the centrality of human factors \cite{faquin2016, diniz2025}.

Agile processes represent modern approaches to software development and project management, prioritizing flexibility, collaboration, rapid delivery, and responsiveness to change. They emerged as an alternative to traditional models driven by rigid plans and sequential processes \cite{Al-Saqqa2020Agile, Abrahamsson2017Agile}. Although each approach emphasizes distinct practices, all require a balanced combination of technical and behavioral skills. The choice of process depends on factors such as project context, team size, and objectives, with Scrum, Kanban, and XP being the most widely adopted in software development environments \cite{Al-Saqqa2020Agile}. Scrum emphasizes the importance of communication to ensure alignment among team members and stakeholders, particularly during daily meetings and sprint reviews \cite{Hidayati2020Hard}. Kanban, which is based on visual workflow management and continuous flow, relies heavily on organizational and time management skills \cite{sampaio2021soft, choque2024assessing}. Similarly, Extreme Programming (XP), which focuses on practices such as pair programming, values teamwork and communication \cite{sampaio2021soft}. The Scaled Agile Framework (SAFe), in turn, targets large organizations by enabling coordinated collaboration among multiple teams, a context in which communication skills are essential for clear information exchange \cite{polakova2023soft}. In the software domain, these approaches reflect adaptive, collaborative, and human-oriented practices aimed at generating value through Information Technology (IT) \cite{lima2025engagement}. Consequently, the development of soft skills is essential for preparing professionals capable of integrating the human and technical dimensions of software development.

Given this context, this study seeks to answer the following research question: \textbf{What are the main soft skills identified in the literature within agile contexts?} Additionally, the study addresses two complementary research questions: \textit{Which soft skills are required for each role within agile development teams?} and \textit{How do different agile methodologies impact the soft skills required in agile development teams?}. These questions aim to broaden the understanding of the relationship between soft skills, roles, and adopted practices, encouraging further investigation into areas related to agile processes and the human aspects of software development. To this end, a Systematic Mapping Study (SMS) was planned and conducted, covering a period of 25 years (from the emergence of the Agile Manifesto to the present). The review analyzed 97 studies, identified 33 distinct soft skills, and documented the main roles in agile teams, their seniority levels, and the agile processes most commonly adopted in industry.

The results of this SMS are intended for different academic and professional audiences, serving the following purposes:

\begin{itemize}
    \item Software Engineering (SE) researchers, by providing a systematized mapping of the most recurrent soft skills in agile contexts, contribute to the consolidation of the theoretical body of knowledge on human factors in software development and support future empirical studies;
    \item Educators and educational institutions, by providing evidence that can support the updating of curricula, courses, and teaching methodologies, fostering the integration of technical and behavioral competencies in the education of computing professionals;
    \item Project managers, technical leaders, and organizational researchers, by enabling reflection on agile team composition, role definition, seniority levels, and the alignment of management practices with the human demands of software development;
    \item Researchers and professionals in the human resources domain, by supporting the development of more well-founded criteria for recruitment, selection, and evaluation processes in agile environments;
    \item Software development organizations and communities of practice, by contributing to the continuous improvement of development processes and promoting more collaborative, adaptive, and human-centered approaches.
\end{itemize}

\section{Related Work} \label{sec:relatedwork}

Matturro et al.~\cite{matturro2019} investigated which soft skills are considered relevant for software engineering practice. As a result, 30 categories of soft skills were identified, including communication (reported in 91\% of the analyzed papers), teamwork (68\%), analytical skills, and organizational/planning skills (both mentioned in 55\% of the studies). In addition, the most commonly used research methods in studies addressing soft skills in software engineering were job advertisements (39\%) and surveys/questionnaires (29\%).

According to Wazlawick et al.~\cite{wazlawick2024}, the growing importance of these skills is driven by the dynamism of the contemporary world and the demands of the labor market. The authors aimed to analyze and synthesize an ontology of soft skills for 21st-century Information Technology (IT) professionals. From an initial selection of 780 articles, 15 were retained, resulting in the identification of 80 unique soft skills. These skills were organized into seven groups based on criteria such as interpersonal versus intrapersonal and basic versus advanced skills. This classification seeks to support pedagogical project design in IT programs, corporate development initiatives, and recruitment processes, while also encouraging the adoption of active learning methodologies to foster more well-rounded professionals.

The study by Mohammed and Ozdamli~\cite{mohammed2024} emphasizes the importance of soft skills in higher education within the IT domain. The authors selected 69 full papers published between 2018 and 2024 that focused on soft skills in higher education. The results revealed that communication, teamwork and collaboration, problem-solving and critical thinking, adaptability and flexibility, leadership, self-awareness and intrapersonal skills, ethics and moral values, creativity and innovation, and socio-emotional skills are crucial for students. The study also identified global trends, such as increased industry involvement in the design of higher education programs to address current demands, and the use of project-based learning as an active methodology to strengthen soft skills.

From the perspective of specific professional roles, Castro et al. \cite{castro2023} highlighted the most relevant soft skills for software analysts, software developers, and data scientists. The research was conducted in two main stages: a literature review of soft skills and a field study involving 21 management professionals with at least 6 years of experience. The outcome was a ranking of the ten most relevant soft skills for each professional profile. The most prominent skills included teamwork, communication, problem-solving, trustworthiness/inspiring confidence, initiative, flexibility, analytical thinking and quality control, coordination and time management, resilience, and creativity.

In the study conducted by Gonçalves et al. \cite{goncalves2024}, the authors investigated the competencies and skills required of computing professionals. A Multivocal Literature Review (MLR) was employed, combined with a survey involving 53 undergraduate students from programs such as Computer Science, Computer Engineering, Information Systems, and Software Engineering. The study identified 14 technical skills and 12 behavioral and social skills, with communication, teamwork, and interpersonal relationships being the most frequently mentioned.

The reviewed studies converge on identifying essential non-technical skills, such as communication, teamwork, and adaptability. This consistency is observed regardless of the methodological approach, including systematic literature reviews (e.g., Matturro et al.~\cite{matturro2019}), ontology-based studies (Wazlawick et al. \cite{wazlawick2024}), investigations in higher education contexts (Mohammed and Ozdamli \cite{mohammed2024}), or analyses focused on specific professional roles (Castro et al. \cite{castro2023}; Gonçalves et al.~\cite{goncalves2024}). However, these studies differ in scope, data sources, and temporal coverage, leading to distinct classifications and groupings of interpersonal skills. In this context, the present study provides a literature review covering a 25-year period (2000–2025), aiming to identify soft skills in agile teams within the industry, resulting in the identification of 33 non-technical skills.

\section{Research Methodology} \label{sec:metodologia}

This study presents a Systematic Mapping Study (SMS) aimed at identifying soft skills in the agile software development industry, covering a retrospective period of 25 years. The research protocol is based on the systematic review process proposed by Petersen et al. \cite{petersen2008systematic, petersen2015guidelines}.

\subsection{Research Questions (RQ)}
The main objective of this research is to conduct an SMS to identify the primary soft skills in the software development industry. To this end, the following Research Questions (RQ) were defined:

\begin{itemize}
    \item[\textbf{RQ$_1$}] \textbf{- What are the main soft skills required in agile software development contexts?} Objective: To build a repository of soft skills that serves as a reference for researchers, educators, and managers in the development of interpersonal competencies in software development environments.
    \item[\textbf{RQ$_2$}] \textbf{- Which soft skills are required for each role within agile development teams?} Objective: To map the specific soft skills required for each role, supporting targeted training strategies and the alignment of professional profiles with their respective responsibilities.
    \item[\textbf{RQ$_3$}] \textbf{- How do different agile methodologies impact the soft skills required in development teams?} Objective: To identify and compare agile methodologies and their variations in fostering soft skills, guiding the adoption of frameworks that best fit team profiles and organizational objectives.
\end{itemize}
    
\subsection{Planning}

Six research databases in the field of Computer Science were selected: ACM Digital Library, IEEE Digital Library Portal, SBC OpenLib (SOL), ScienceDirect, Scopus, and Springer Link. To guide the study, the Population, Intervention, Comparison, Outcome, Context (PICOC) methodology was applied \cite{liberati2009} (Table \ref{tab:picoc}).

\begin{table}[!h]  \small
\centering
\caption{PICOC protocol used in the SMS}
\label{tab:picoc}
\begin{tabular}{p{0.2\linewidth} p{0.7\linewidth}}
\hline
\multicolumn{2}{c}{\textbf{PICOC Protocol}} \\ \hline
\textbf{Population}   & Computing professionals \\ \hline
\textbf{Intervention} & Participation in agile teams \\ \hline
\textbf{Comparison}   & Not applicable \\ \hline
\textbf{Outcomes}     & Identification and categorization of required soft skills \\ \hline
\textbf{Context}      & Software development using agile methodologies \\ \hline
\end{tabular}
\end{table}

The “Comparison” element was defined as “not applicable” due to the nature of the study, as this work does not aim to compare interventions, methods, or distinct groups, but rather to identify and categorize the interpersonal skills required in agile teams based on the existing literature. Therefore, the focus is on the synthesis and organization of knowledge, rather than on the comparative analysis of approaches.

The construction of the search string was based on the identification of the elements of the PICOC strategy. The search string (Table \ref{tab:searchstring}) was translated into Portuguese for searches conducted in the SBC OpenLib (SOL) database. To ensure consistency in data extraction, a structured data extraction form was developed according to the RQ. This form specified which information should be collected from each article, including identification of soft skills, associated roles, agile methodologies mentioned, seniority levels, and observed trends. Table \ref{tab:planodeextracao} presents the mapping of the RQs in relation to the extracted data, serving as a guide for the next phase of the SMS.

\begin{table}[!h]  \small
\centering
\scriptsize
\caption{Search String}
\label{tab:searchstring}
\begin{tabular}{p{0.2\linewidth} p{0.7\linewidth}}
\hline
\textbf{Language} & \textbf{Search String} \\ \hline

\textbf{English} &
("information technology" OR "computer") AND
("software development" OR "software industry") AND
("agile" OR "scrum") AND
("soft skills" OR "interpersonal skills" OR "non-technical skills" OR
"human factor" OR "social skills" OR "people skills" OR
"personal skills" OR "non-cognitive skills")
\\ \hline

\textbf{Portuguese} &
("tecnologia da informação" OR "computação") AND
("desenvolvimento de software" OR "indústria de software") AND
("ágil" OR "scrum") AND
("soft skills" OR "habilidades interpessoais" OR
"habilidades não técnicas" OR "fator humano" OR
"habilidades sociais" OR "habilidades com pessoas" OR
"habilidades pessoais" OR "habilidades não cognitivas")
\\ \hline
\normalsize
\end{tabular}
\end{table}

\begin{table}[!h] \small
\centering
\scriptsize
\caption{Data Extraction Plan}
\label{tab:planodeextracao}
\begin{tabular}{p{0.3\linewidth} p{0.58\linewidth}}
\hline
\textbf{Element} & \textbf{Description} \\ \hline

\textbf{Research Question} & RQ1 – Main required soft skills \\ \hline
\textbf{Extracted Data} & List of identified soft skills \\ \hline
\textbf{Analysis Approach} & Frequency distribution \\ \hline
\hline

\textbf{Research Question} & RQ2 – Soft skills by development team role \\ \hline
\textbf{Extracted Data} & Role associated with soft skills \\ \hline
\textbf{Analysis Approach} & Role-based mapping \\ \hline
\hline

\textbf{Research Question} & RQ3 – Impact of agile methodologies \\ \hline
\textbf{Extracted Data} & Agile methodology used \\ \hline
\textbf{Analysis Approach} & Comparative analysis across methodologies \\ \hline
\normalsize
\end{tabular}
\end{table}

\subsection{Inclusion and Exclusion Criteria}
The inclusion and exclusion criteria were defined to identify primary studies relevant to this research. The establishment of these criteria enables the selection of retrieved studies with the highest potential to answer the research questions (RQs). Table \ref{tab:criterios} presents the defined criteria.

The inclusion of studies published in Portuguese and Spanish was established as an explicit criterion of this research, rather than an arbitrary choice. This decision was motivated by the relevance of contributions from Latin America and the Iberian Peninsula to Software Engineering, particularly in contexts that are often underrepresented in English-dominated literature. As this study aims to analyze a specific regional context, other languages, such as German, French, or Chinese, were not included due to scope limitations and the lack of direct alignment with the regional focus of this study.

\begin{table}[!h]
\scriptsize
\centering
\caption{Inclusion and Exclusion Criteria of the SMS}
\label{tab:criterios}
\begin{tabular}{p{0.1\linewidth} p{0.7\linewidth}}
\toprule

\multicolumn{2}{c}{\textbf{Inclusion Criteria (IC)}} \\ \midrule
\textbf{IC.1} & Primary studies (original research). \\ \hline
\textbf{IC.2} & Studies conducted between 2000 and May 2025. \\ \hline
\textbf{IC.3} & Studies written in Portuguese, English, and/or Spanish. \\ \hline
\textbf{IC.4} & Studies addressing soft skills in software development teams or roles. \\ \midrule

\multicolumn{2}{c}{\textbf{Exclusion Criteria (EC)}} \\ \midrule
\textbf{EC.1} & Secondary studies, such as literature reviews, mappings, tertiary studies, and other reviews. \\ \hline
\textbf{EC.2} & Studies not available for free download or institutional access. \\ \hline
\textbf{EC.3} & Studies focused on computing education or student perspectives. \\ \hline
\textbf{EC.4} & Studies not peer-reviewed, such as book chapters, technical reports, documents, abstracts, or presentations. \\ \hline
\textbf{EC.5} & Studies that do not address soft skills in software development teams or roles. \\

\bottomrule
\end{tabular}
\end{table}
\normalsize

\subsection{Conduction}
\begin{figure*}[!h]
\centering
\includegraphics[width=0.8\textwidth, keepaspectratio]{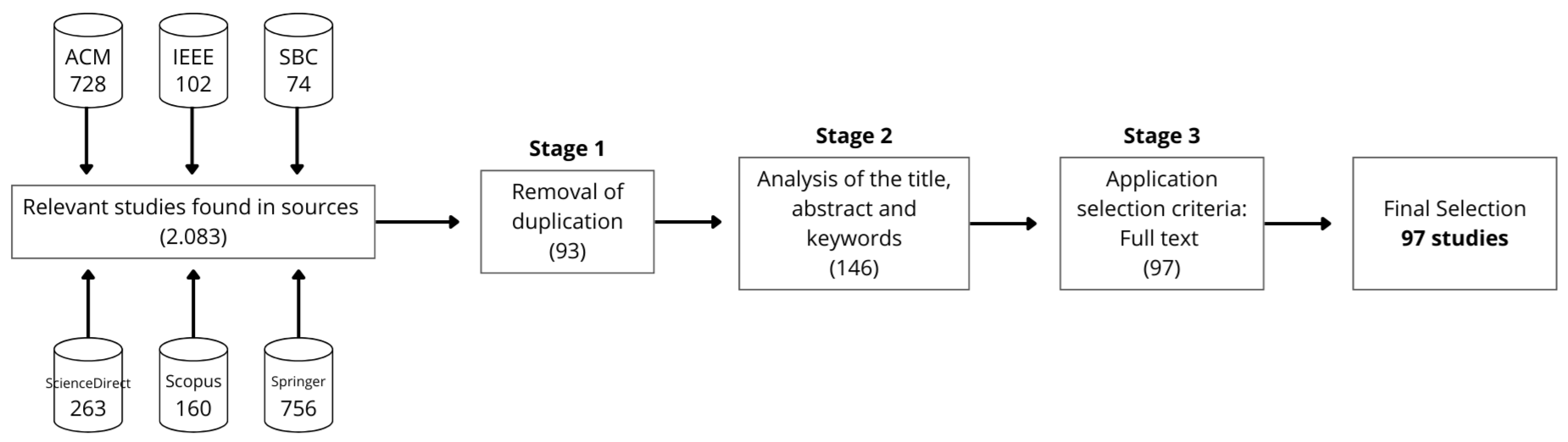}
\caption{Study search and selection process.}
\label{fig:metodologia}
\end{figure*}

The conduction phase of the SMS corresponds to the execution of the planning stage. Due to the large number of studies retrieved from the databases (2,083 articles), the selection process was divided into three stages, as illustrated in Figure \ref{fig:metodologia}: (i) Duplicate identification: all studies extracted from the databases were checked for duplication using the online tool Parsifal. As a result, 93 duplicate records were removed. Next, (ii) Title and abstract screening: the 1,146 studies remaining from the previous stage were assessed for relevance to the research questions, considering titles, abstracts, and keywords. Studies that did not meet the inclusion criteria or fell under the exclusion criteria (Table \ref{tab:criterios}) were discarded. Finally, (iii) Full-text reading: studies identified as relevant were analyzed in full to ensure that only those effectively addressing soft skills in software development teams or roles were included in the final synthesis.

The overall selection process is presented in Figure \ref{fig:metodologia}, following the PRISMA model, which highlights the number of studies identified, excluded, duplicated, and accepted at each stage (Table \ref{tab:basededados2}). Furthermore, the conduction and selection of studies were performed by authors with complementary expertise: a PhD professor in Computer Science specializing in human aspects of Software Engineering (SE), and a master’s student in Software Engineering focused on human aspects in SE. The 97 selected studies are available in \cite{repositorio}, each identified with the prefix “S” (for Study), followed by a chronological identification number.

\begin{table}[!h]
\centering
\scriptsize
\caption{Search results across databases}
\label{tab:basededados2}

\begin{tabular}{p{0.15\linewidth}
                p{0.12\linewidth}
                p{0.12\linewidth}
                p{0.12\linewidth}
                p{0.12\linewidth}
                p{0.12\linewidth}}
\hline
\textbf{Database} &
\textbf{Identified} &
\textbf{Duplicates} &
\textbf{Filter 1} &
\textbf{Filter 2} &
\textbf{Accepted} \\ \hline

ACM Digital Library  & 728 & 17 & 42 & 25 & 25 \\ 
IEEE Digital Library & 102 & 6  & 30 & 10 & 10 \\ 
SBC OpenLib (SOL)    & 74  & 5  & 20 & 18 & 18 \\ 
ScienceDirect        & 263 & 5  & 15 & 12 & 12 \\ 
Scopus               & 160 & 25 & 29 & 22 & 22 \\ 
Springer             & 756 & 35 & 10 & 10 & 10 \\ \hline

\textbf{Total}       & \textbf{2,083} & \textbf{93} & \textbf{146} & \textbf{97} & \textbf{97} \\ \hline

\end{tabular}
\end{table}

\subsection{Data Extraction and Analysis}
To answer the RQs defined in the planning phase, a detailed reading and analysis of the 97 articles obtained in the review process were conducted. For each selected study, relevant elements were recorded to address the RQ (as presented in Table \ref{tab:planodeextracao}). In addition to general publication information (title, date, authors, and publication venue), the extraction form included the following aspects: (i) identification and definition of the soft skills reported in the study; (ii) association of soft skills with specific roles or general team analyses; and (iii) indication of the agile methodologies present in the research context and the related soft skills.

The definitions found in the studies supported standardizing the understanding of non-technical skills. In many cases, the same interpersonal competence was named differently across studies. To address this, soft skills were consolidated based on the described activities or conceptual meanings to reduce redundancy and ensure consistent categorization. After extraction, the data were classified into analytical categories that encompassed the distribution of soft skills by role, seniority level, and agile methodology. Data organization was performed using Google Sheets, described in the results section and available at \cite{repositorio}.

\section{Results} \label{sec:resultados}

\begin{figure}[h]
\centering
\includegraphics[scale=0.4]{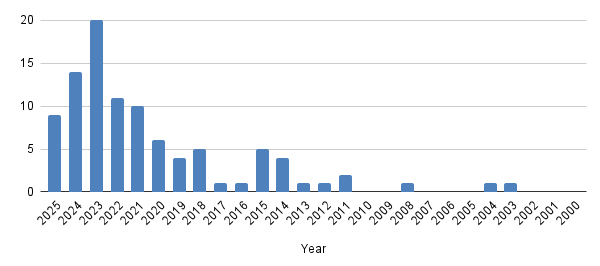}
\caption{Frequency per year}
\label{fig:anografico}
\end{figure}

This section presents the main findings of the SLR, based on the 97 analyzed studies, available at \cite{repositorio}. Figure \ref{fig:anografico} shows the distribution of studies over the last 25 years. During the COVID-19 pandemic period (2020–2023), 47 studies were published, reflecting an increased research interest in examining the sudden and dynamic context experienced by computing professionals.

\subsection{RQ1 - What are the main soft skills required in agile development contexts?}

The analysis of the 97 studies included in the SLR resulted in the identification of 33 distinct soft skills. It was observed that several soft skills were named differently across studies. Based on this observation, the skills were grouped according to similar activities or equivalent conceptual meanings. Among them, a smaller subset appears repeatedly across multiple studies, suggesting a core set of highly valued competencies in agile environments. These include: Communication, Teamwork, Leadership, Adaptability, Proactivity, Motivation, Problem Solving, Organization, Creativity, Responsibility, Socialization, Willingness to Learn, Trust, Knowledge Management, Business Vision, Empathy, Critical Thinking, Negotiation, Active Listening, Ethics, Conflict Management, and Goal Setting. The data are presented in Table \ref{tab:softskillsencontradas}, which lists the 22 main interpersonal skills (the remaining skills and their definitions are available at \cite{repositorio}).

Additionally, the trends of soft skills were analyzed based on their temporal evolution over the evaluated period. Each skill was examined according to its citation frequency across studies over the years, enabling the identification of patterns of growth, stability, or decline. This approach allowed the classification of competencies as emerging, consolidated, or declining, providing a longitudinal perspective on the evolution of their relevance in the literature.

\textbf{Communication} is the most frequently mentioned skill, with 87 occurrences (89.6\%), and is considered essential for ensuring stakeholder alignment, clarity in requirements communication, and efficiency in everyday team interactions. The second most cited skill is \textbf{Teamwork} (63 mentions, 64.9\%), reflecting the central role of collaboration in agile methodologies. The ability to cooperate, negotiate, and share responsibilities is crucial for teams to achieve collective goals through iterations and increments. \textbf{Leadership} (45 mentions, 46.3\%) also stands out, but as a distributed competency among team members who assume roles of facilitation, motivation, and decision-making. This reinforces the concept of shared leadership, which is typical of self-organizing teams. \textbf{Adaptability} (36 mentions, 37.1\%) emerges as a key attribute in the face of rapid market and technological changes. \textbf{Proactivity} (32 mentions, 32.9\%) refers to the practice of identifying and initiating tasks rather than merely reacting to previously assigned demands. \textbf{Motivation} (29 mentions, 29.9\%), expressed through enthusiasm, engagement, and task focus, is essential for the development of successful projects.

\textbf{Problem Solving} (29 mentions, 29.9\%) supports overcoming complex challenges through informed and rational decision-making. In software development, this skill is often associated with debugging activities aimed at investigating and identifying the causes of software errors or failures. \textbf{Creativity} (26 mentions, 26.8\%) corresponds to the ability to approach problems from different perspectives and propose innovative solutions. In software engineering, this extends beyond design and encompasses the resolution of complex technical challenges. \textbf{Organization} (24 mentions, 24.7\%) relates to the ability to manage tasks efficiently, ensuring timely delivery and appropriate resource utilization. In development contexts, this competence can be observed in practices such as version control and source code management. Similarly, \textbf{Willingness to Learn} (23 mentions, 23.7\%) directly impacts the rapid acquisition of new concepts, methodologies, and technologies, which is crucial in the constantly evolving field of computing.

\textbf{Trust} (23 mentions, 23.7\%) is related to team members’ profiles, requiring competence, reliability, and commitment to project goals. With 20 mentions (20.6\%), \textbf{Responsibility} refers to reliability and consistent delivery of results. In agile teams, responsibility tends to be shared, meaning that commitment to sprint goals is collectively assumed. \textbf{Empathy} (19 mentions, 19.5\%) refers to the ability to understand and place oneself in the position of users, clients, or team members. Usability testing and user-centered design are examples of this competence. In addition, \textbf{Socialization} appears as a skill with 18 mentions (18.5\%), being fundamental in development environments, especially in agile teams that value interaction. To understand business domains, including backlog prioritization and requirements analysis, \textbf{Business Vision} (18 mentions, 18.5\%) aligns technical decisions with organizational strategic objectives and end-user needs.

\textbf{Critical Thinking} (17 mentions, 17.5\%) enables objective analysis of information, identification of alternative solutions, and informed decision-making; code refactoring and requirements analysis are examples of its application. Skills such as \textbf{Negotiation} and \textbf{Knowledge Management} each received 16 mentions (16.4\%). Negotiation involves planning interactions (e.g., Scrum sprints) and scope definition, focusing on reaching agreements or resolving divergences within the development team. Knowledge Management refers to the ability to share information and experiences effectively within the team, such as mentoring new members and documenting code and processes.

\textbf{Ethics} (15 mentions, 15.5\%) in software development relates to acting with integrity and responsibility, considering the social impact and consequences of developed technologies. In identifying, addressing, and resolving divergences, \textbf{Conflict Management} (13 mentions, 13.4\%) plays a direct role in discussions among team members, which may involve conflicting opinions regarding technologies or expectation alignment. \textbf{Active Listening} (13 mentions, 13.4\%) involves attentive listening to understand shared information and is essential during requirements meetings, technical discussions, and feedback sessions. Defining measurable and achievable goals, both individually and collectively, is fundamental to maintaining focus during sprints, highlighting the importance of \textbf{Goal Setting} (11 mentions, 15\%).

Other competencies include: \textbf{Productivity}, with 3 mentions (3.1\%); \textbf{Working Under Pressure}, with 5 mentions (5.2\%); \textbf{Analytical Skills}, \textbf{Curiosity}, \textbf{Ability to Give and Receive Feedback}, and \textbf{Patience}, all with 6 mentions (6.2\%); \textbf{Time Management}, with 8 mentions (8.2\%); \textbf{Quality Orientation} and \textbf{Innovation}, both with 9 mentions (9.3\%); and finally, \textbf{Emotional Intelligence} and \textbf{Self-Management}, both with 10 mentions (10.3\%).

Based on the analysis of the publications, non-technical skills were classified into two occurrence patterns over the years: \textit{emerging} and \textit{stable} skills. This categorization (also presented in Table \ref{tab:softskillsencontradas}) provides an overview of the evolution of competencies over 25 years, considering trends and market demands. Among emerging skills, consistent growth is observed in competencies such as \textbf{Communication}, \textbf{Adaptability}, \textbf{Problem Solving}, \textbf{Willingness to Learn}, \textbf{Knowledge Management}, and \textbf{Critical Thinking}. This trend is discussed in studies such as Wong et al. \cite{wong_its_2025} and Hamza et al. \cite{luong_agile_2021}, which highlight rapid transformations in today’s market, requiring more flexible and collaborative approaches, such as agile methodologies that place human factors at the center of project success.

Stable skills correspond to competencies that maintained a relatively constant frequency of mention throughout the analyzed period, including \textbf{Teamwork}, \textbf{Leadership}, \textbf{Proactivity}, \textbf{Motivation}, \textbf{Organization}, \textbf{Creativity}, \textbf{Responsibility}, \textbf{Socialization}, \textbf{Trust}, \textbf{Business Vision}, \textbf{Empathy}, \textbf{Negotiation}, \textbf{Active Listening}, \textbf{Ethics}, \textbf{Conflict Management}, and \textbf{Goal Setting}, suggesting that their importance has remained well established in the literature over the past 25 years. However, this does not imply a decrease in relevance; rather, these competencies have become fundamental and are considered basic prerequisites in modern software development environments, especially with the widespread adoption of agile methodologies \cite{mnkandla_balancing_nodate, mangiza_requisite_2020, assyne_essential_2022, hoda_self-organizing_2013, coelho2024, fernandez2011influence}.

\begin{table*}[!h]
\scriptsize
\centering
\caption{Identified soft skills}
\label{tab:softskillsencontradas}
\begin{tabular}{c c p{2cm} p{5cm} p{1cm} p{7cm}}
\hline
\textbf{Trend} & \textbf{ID} & \textbf{Soft Skill} & \textbf{Definition} & \textbf{Mentions} & \textbf{Studies (97)} \\ \hline

$\blacktriangle$ & 1 & Communication &
Ability to convey information clearly and effectively, whether orally or in writing. &
87 (89.6\%) &
[S1]–[S10], [S12]–[S13], [S15]–[S24], [S26]–[S29], [S31]–[S36], [S37]–[S40], [S42]–[S45], [S46]–[S53], [S54], [S56]–[S58], [S60]–[S62], [S64]–[S66], [S67]–[S69], [S71]–[S82], [S84]–[S97] \\ \hline

$\blacksquare$ & 2 & Teamwork &
Ability to work effectively in a collaborative environment to achieve common goals. &
63 (64.9\%) &
[S4], [S7]–[S17], [S19], [S20], [S22], [S24], [S25], [S27], [S28], [S30]–[S32], [S34], [S36], [S37], [S39], [S40], [S42]–[S45], [S47]–[S50], [S52], [S54]–[S58], [S62], [S64]–[S66], [S69], [S71], [S72], [S74], [S76]–[S78], [S80]–[S82], [S84]–[S86], [S88], [S92]–[S93], [S95]–[S96] \\ \hline

$\blacksquare$ & 3 & Leadership &
Ability to inspire, motivate, and guide individuals or teams, promoting high performance. &
45 (46.3\%) &
[S6], [S8]–[S10], [S13], [S15], [S16], [S18], [S19], [S22]–[S24], [S27]–[S29], [S32], [S37], [S42]–[S44], [S48]–[S50], [S54], [S58], [S60], [S62], [S64]–[S67], [S71], [S75]–[S80], [S83]–[S85], [S92]–[S94], [S96] \\ \hline

$\blacktriangle$ & 4 & Adaptability &
Ability to quickly adjust to changes in the environment, requirements, or processes. &
36 (37.1\%) &
[S1], [S7], [S9], [S11], [S15]–[S18], [S20], [S23]–[S25], [S28]–[S29], [S36], [S38], [S40], [S47], [S49], [S52], [S59], [S60], [S64], [S65], [S67], [S70], [S73], [S76]–[S78], [S80]–[S81], [S85], [S91], [S92], and [S97] \\ \hline

$\blacksquare$ & 5 & Proactivity &
Ability to anticipate demands and act independently to solve problems. &
32 (32.9\%) &
[S8], [S10], [S20], [S25], [S30], [S35], [S37], [S40], [S43], [S45], [S47]–[S50], [S52], [S55], [S58]–[S60], [S62], [S64], [S66], [S70], [S76], [S78], [S81], [S84], [S85], [S87], [S88], [S91], [S93] \\ \hline

$\blacksquare$ & 6 & Motivation &
Engagement, enthusiasm, and focus on task performance. &
29 (29.9\%) &
[S2], [S4], [S12]–[S13], [S15], [S17], [S18], [S20], [S28]–[S30], [S36]–[S38], [S43], [S49], [S50], [S59], [S60], [S64]–[S65], [S71], [S76], [S77], [S81], and [S83]–[S86] \\ \hline

$\blacktriangle$ & 7 & Problem Solving &
Ability to understand, analyze, and solve complex problems. &
29 (29.9\%) &
[S7], [S13], [S16], [S18]–[S20], [S23], [S24], [S30]–[S32], [S35], [S38], [S43], [S46], [S49], [S50], [S52], [S53], [S58], [S64], [S69], [S78]–[S80], [S82], [S84], [S85], [S92] \\ \hline

$\blacksquare$ & 8 & Creativity &
Ability to propose innovative solutions and approach problems from new perspectives. &
26 (26.8\%) &
[S14], [S16], [S23]–[S25], [S31]–[S33], [S38], [S44]–[S47], [S49], [S50], [S54], [S58], [S60], [S62], [S65], [S66], [S69], [S78], [S85], [S86], [S88] \\ \hline

$\blacksquare$ & 9 & Organization &
Ability to manage tasks, time, and priorities efficiently. &
24 (24.7\%) &
[S1], [S2], [S13], [S14], [S19], [S20], [S23], [S29], [S35], [S39], [S43], [S47], [S49], [S50], [S53], [S65]–[S67], [S78], [S81], [S84], [S85], [S87], [S93] \\ \hline

$\blacksquare$ & 10 & Trust &
Willingness to trust and be trustworthy in professional relationships. &
23 (23.7\%) &
[S3], [S7], [S9], [S11]–[S12], [S14], [S16], [S18], [S26], [S36], [S45], [S51], [S52], [S61], [S64], [S65], [S70], [S71], [S74], [S77], [S78], [S93], [S97] \\ \hline

$\blacktriangle$ & 11 & Willingness to Learn &
Ability to rapidly acquire new knowledge and skills. &
23 (23.7\%) &
[S7], [S10], [S16], [S20], [S22], [S24], [S35], [S43], [S49]–[S50], [S52], [S54], [S60], [S62], [S64], [S70], [S71], [S76], [S78], [S81], [S84], [S91], [S92] \\ \hline

$\blacksquare$ & 12 & Responsibility &
Commitment to reliable and high-quality deliverables. &
20 (20.6\%) &
[S7], [S11], [S13], [S20], [S23], [S31], [S35]–[S38], [S45], [S49], [S64], [S66], [S67], [S76], [S77], [S81], [S84], and [S93] \\ \hline

$\blacksquare$ & 13 & Empathy &
Ability to understand others’ perspectives and emotions. &
19 (19.5\%) &
[S6], [S23], [S52], [S56], [S58], [S60], [S64], [S65], [S67], [S68], [S70], [S80], [S83], [S84], [S85], [S89], [S93], [S94], [S95] \\ \hline

$\blacksquare$ & 14 & Socialization &
Ability to engage socially and integrate with the team. &
18 (18.5\%) &
[S4], [S6], [S7], [S9], [S12], [S13], [S21], [S23], [S25], [S28], [S36], [S44], [S49], [S55], [S81], [S84], [S90] \\ \hline

$\blacksquare$ & 15 & Business Vision &
Understanding of organizational context and business needs. &
18 (18.5\%) &
[S6], [S9], [S13], [S15], [S19], [S20], [S22], [S23], [S32], [S33], [S36], [S52], [S57], [S71], [S77], [S81], [S82], [S84] \\ \hline

$\blacktriangle$ & 16 & Critical Thinking &
Ability to analyze information logically and reflectively. &
17 (17.5\%) &
[S16], [S20], [S22], [S23], [S27], [S32], [S38], [S49], [S54], [S65]–[S67], [S78], [S81], [S83], [S85], [S91] \\ \hline

$\blacktriangle$ & 17 & Knowledge Management &
Ability to share, document, and reuse knowledge. &
16 (16.4\%) &
[S3], [S9], [S26], [S28], [S34], [S44], [S50], [S57], [S61], [S79], [S84], [S85], [S89], [S91], [S93], [S96] \\ \hline

$\blacksquare$ & 18 & Negotiation &
Ability to reach agreements through dialogue and concessions. &
16 (16.4\%) &
[S16], [S18], [S19], [S23], [S39], [S58], [S60], [S65]–[S68], [S78], [S79], [S85], [S89], [S93] \\ \hline

$\blacksquare$ & 19 & Ethics &
Professional behavior based on integrity and social responsibility. &
15 (15.4\%) &
[S4], [S20], [S22], [S23], [S39], [S49], [S52], [S62], [S64]–[S66], [S70], [S81], [S93], [S96] \\ \hline

$\blacksquare$ & 20 & Conflict Management &
Ability to deal with conflicts constructively. &
13 (13.4\%) &
[S15], [S18], [S23], [S28], [S39], [S49], [S51], [S62], [S65], [S67], [S68], [S75], [S93] \\ \hline

$\blacksquare$ & 21 & Active Listening &
Ability to listen attentively and consider different viewpoints. &
13 (13.4\%) &
[S3], [S10], [S28], [S32], [S44], [S49], [S58], [S60], [S64], [S78], [S84], [S89], [S93] \\ \hline

$\blacksquare$ & 22 & Goal Setting &
Establishment of clear and measurable objectives. &
11 (11.3\%) &
[S13], [S18], [S19], [S23], [S28], [S39], [S49], [S82], [S84], [S85], [S93] \\ \hline

\end{tabular}

\vspace{0.3em}
\textbf{Legend:} $\blacktriangle$ increasing trend; $\blacksquare$ stable trend
\end{table*}
\normalsize

\subsection{RQ2 - Which soft skills are required for each role within agile development teams?}

To complement the study, a research question was formulated with the aim of identifying the skills associated with each role (Table \ref{tab:softskillpapel}). The roles were extracted directly from the review stage, while their definitions were based on previous studies \cite{lenberg2015, shastri_understanding_2017, gomes2020, rialti_leaders_2024, michael_ayas_roles_2024, silva_farias_what_2025}. Although the categorization reflects the roles reported in the analyzed studies, some of them do not explicitly associate soft skills with specific roles, which explains the lower number of mentions observed in the heatmap. Regarding the distribution of roles, there is a clear predominance of the \textbf{Developer} role, with 62 mentions (63.9\%), reinforcing its centrality in the analyzed contexts. This is followed by \textbf{Technical Leaders} (25; 25.\%), and \textbf{Requirements Analysts} and \textbf{Product Owners} (19; 19.6\% each). The roles of \textbf{Product Manager} (16; 16.5\%), \textbf{Designers} (15; 15.5\%), and the \textbf{Team as a whole} (12; 12.4\%) also appear with relevant frequency. In contrast, more specialized roles, such as \textbf{Software Architect} (8; 8.2\%), \textbf{Data Scientist} (3; 3.1\%), \textbf{Security Analyst} and \textbf{Support Analysts} (2; 2.1\% each), and \textbf{DevRel} (1; 1.0\%), appear less frequently in the analyzed literature.

Across roles, \textbf{Communication} emerges as the most recurrent competency, with high incidence among \textbf{Developers} (55 mentions), \textbf{Scrum Masters} and \textbf{QA/Test Analysts} (28 each), \textbf{Technical Leaders} (22), and \textbf{Requirements Analysts} (18). Similarly, \textbf{Teamwork} shows a wide distribution, standing out among \textbf{Developers} (41), \textbf{Scrum Masters} (25), \textbf{Technical Leaders} (16), and \textbf{Requirements Analysts} (11), reinforcing its importance in collaborative and coordination-oriented roles.

The competencies \textbf{Leadership} and \textbf{Adaptability} present a more contextual distribution. Leadership is mainly observed among \textbf{Developers} (26) and \textbf{Technical Leaders} (16), but also appears among \textbf{Product Owners} and \textbf{Designers} (9 each), suggesting a distributed nature in agile environments. Adaptability, in turn, is concentrated among \textbf{Developers} (26), \textbf{Scrum Masters} (16), \textbf{QA/Test Analysts} (15), and \textbf{Technical Leaders} (13), reflecting the dynamic nature of these roles.

Behavioral competencies such as \textbf{Proactivity} and \textbf{Motivation} are concentrated among \textbf{Developers} (21 in both cases), but also appear among \textbf{Scrum Masters} (16 and 12) and \textbf{QA/Test Analysts} (13 and 8). \textbf{Problem Solving}, in turn, presents a more balanced distribution, with emphasis on \textbf{Developers} (20), \textbf{QA/Test Analysts} (15), and \textbf{Scrum Masters} (12). Less frequent competencies remain relevant in specific contexts. \textbf{Organization} stands out among \textbf{Developers} (15) and \textbf{Scrum Masters} (8), while \textbf{Creativity} appears mainly among \textbf{Developers} and \textbf{QA/Test Analysts} (15 each). \textbf{Responsibility} is more associated with \textbf{QA/Test Analysts} (11), \textbf{Scrum Masters} (10), and \textbf{Technical Leaders} (7).

Other competencies, such as \textbf{Socialization} (\textbf{Developers}: 11; \textbf{QA/Test Analysts}: 8; \textbf{Scrum Masters}: 7) and \textbf{Willingness to Learn} (\textbf{Developers}: 14; \textbf{Scrum Masters} and \textbf{QA/Test Analysts}: 9 each), highlight the importance of interaction and continuous learning. Likewise, \textbf{Trust} is distributed among \textbf{Developers} (15), \textbf{Scrum Masters} (11), and \textbf{Product Owners} (7), while \textbf{Empathy}, \textbf{Business Vision}, and \textbf{Knowledge Management} appear mainly among \textbf{Developers} (13, 11, and 10, respectively) and \textbf{Technical Leaders} (4–6).

Finally, competencies related to advanced communication, such as \textbf{Active Listening} (\textbf{Developers}: 10; \textbf{Technical Leaders}: 7) and \textbf{Negotiation} (\textbf{Developers}: 12; \textbf{Technical Leaders}: 4), reinforce the importance of alignment and conflict management. Competencies such as \textbf{Ethics} (\textbf{Developers}: 11; \textbf{Scrum Masters}: 8), \textbf{Conflict Management} (\textbf{Developers}: 10; \textbf{Technical Leaders}: 6), and \textbf{Goal Setting} (\textbf{Developers}: 8; \textbf{Scrum Masters}: 6) appear less frequently, but remain relevant.

Overall, the results indicate that, although different roles exhibit distinct patterns of socio-emotional competencies, there is a central core of skills—particularly \textbf{Communication} and \textbf{Teamwork}—shared across roles, alongside more specific competencies that vary according to the responsibilities of each role.

\begin{table*}[!h] \small
\centering
\caption{Soft Skills by Role}
\label{tab:softskillpapel}
\begin{tabular}{p{2cm} c c c c c c c c c c}
\toprule
\textbf{Soft Skill} &
\textbf{\small{\shortstack{Requirements\\Analyst}}} &
\textbf{\small{Developer}} &
\textbf{\small{Designer}} &
\textbf{\small{\shortstack{Entire\\Team}}} &
\textbf{\small{\shortstack{Technical\\Leaders}}} &
\textbf{\small{\shortstack{Scrum\\Master}}} &
\textbf{\small{\shortstack{QA/Testing\\Analyst}}} &
\textbf{\small{\shortstack{Software\\Architect}}} &
\textbf{\small{\shortstack{Product\\Manager}}} &
\textbf{\small{\shortstack{Product\\Owner}}} \\
\midrule
Communication 
& 18 & 55 & 13 & 12 & 22 & 28 & 28 & 7 & 13 & 17 \\ \hline

Teamwork 
& 11 & 41 & 8 & 8 & 16 & 25 & 18 & 6 & 10 & 11 \\ \hline

Leadership 
& 8 & 26 & 9 & 7 & 16 & 6 & 12 & 5 & 8 & 9 \\ \hline

Adaptability 
& 8 & 26 & 5 & 2 & 13 & 16 & 15 & 4 & 6 & 10 \\ \hline

Proactivity 
& 5 & 21 & 2 & 2 & 10 & 16 & 13 & 4 & 5 & 5 \\ \hline

Problem Solving 
& 8 & 20 & 6 & 2 & 7 & 12 & 15 & 4 & 4 & 5 \\ \hline

Motivation 
& 10 & 21 & 5 & 3 & 10 & 12 & 8 & 4 & 5 & 9 \\ \hline

Creativity 
& 5 & 15 & 5 & 1 & 5 & 9 & 15 & 1 & 5 & 5 \\ \hline

Organization 
& 7 & 15 & 1 & 1 & 5 & 8 & 7 & 3 & 2 & 6 \\ \hline

Trust 
& 1 & 15 & 2 & 3 & 3 & 11 & 6 & 3 & 2 & 7 \\ \hline

Willingness to Learn 
& 5 & 14 & 3 & 4 & 5 & 9 & 9 & 3 & 2 & 3 \\ \hline

Responsibility 
& 5 & 14 & 1 & 1 & 7 & 10 & 11 & 3 & 1 & 5 \\ \hline

Empathy 
& 1 & 13 & 2 & 3 & 4 & 12 & 6 & 3 & 1 & 2 \\ \hline

Business Acumen 
& 3 & 11 & 2 & 4 & 6 & 6 & 4 & 3 & 1 & 3 \\ \hline

Socialization 
& 6 & 11 & 4 & 2 & 6 & 7 & 8 & 2 & 4 & 4 \\ \hline

Critical Thinking 
& 4 & 10 & 5 & 2 & 6 & 7 & 9 & 2 & 1 & 4 \\ \hline

Knowledge Management 
& 4 & 10 & 4 & 2 & 6 & 6 & 6 & 3 & 3 & 5 \\ \hline

Negotiation 
& 1 & 12 & 3 & 0 & 4 & 6 & 9 & 1 & 2 & 5 \\ \hline

Ethics 
& 4 & 11 & 2 & 0 & 2 & 8 & 6 & 2 & 0 & 3 \\ \hline

Active Listening 
& 3 & 10 & 4 & 1 & 7 & 6 & 8 & 2 & 3 & 2 \\ \hline

Conflict Management 
& 3 & 10 & 4 & 2 & 6 & 4 & 5 & 0 & 3 & 4 \\ \hline

Goal Setting 
& 3 & 8 & 2 & 0 & 3 & 6 & 3 & 2 & 2 & 3 \\ 

\bottomrule
\normalsize

\end{tabular}
\end{table*}

\subsection{RQ3 - How do different agile methodologies impact the soft skills required in agile development teams?}

Regarding agile methodologies, Scrum is the most recurrent approach, appearing in more than half of the studies (55 mentions). Extreme Programming (XP) appears with a frequency of 13 mentions, generally associated with Scrum in studies that discuss combined practices or compare approaches. When combined, methodologies such as Scrum and XP appear together in one study. Kanban and SAFe are rarely mentioned (only one mention each), suggesting a gap in the research landscape. In addition, 41 studies do not specify the agile methodology used, representing a significant portion of the analyzed works.

Table \ref{tab:softskillsmetodologias} explores the relationship between methodologies, activities, and soft skills. Not all articles explicitly establish this relationship. \textbf{Scrum}, as a project management–focused methodology with defined roles (Scrum Master, Product Owner, Development Team) and regular ceremonies (Sprint Planning, Daily Scrum, Sprint Review, and Retrospective), intensifies the need for Communication, Active Listening, and Socialization. The adoption of Scrum ceremonies proved to be a relevant factor in streamlining communication, reducing rework, and improving collaborative decision-making. In contrast, \textbf{eXtreme Programming (XP)} emphasizes technical practices and intense collaboration, such as pair programming, requiring teamwork, communication, patience, and constructive feedback skills. \textbf{SAFe}, in turn, emphasizes soft skills related to knowledge management and teamwork to support the integration of specialized knowledge across teams. 

Although all agile methodologies acknowledge the importance of a common core of skills—such as communication and teamwork—each methodology, through its specific practices, emphasizes particular soft skills. Thus, agile methodologies, especially Scrum, not only influence but also reinforce soft skills related to social interaction, adaptability, and collaboration.

\begin{table}[h]  \small

\centering
\caption{Soft Skills and Agile Methodologies} \label{tab:softskillsmetodologias}
\begin{tabular}{p{1cm}p{2cm}p{2cm}p{2cm}}
\hline
\scriptsize
\textbf{Agile Methodology} & \textbf{Activity} & \textbf{Associated Soft Skills} & \textbf{Studies} \\ \hline
Scrum & Daily Scrum, Sprint Planning, Retrospectives & Communication, Active Listening and Socialization & [S14], [S20], [S21], [S22], [S79], [S86], [S97] \\ \hline
XP & Pair Programming & Communication, Teamwork, and Productivity & [S1], [S15], [S32], [S39], [S59] \\ \hline
Scrum and XP & Meetings and Interactions & Communication and Active Listening & [S12], [S23], [S28], [S56], [S79]  \\ \hline
SAFe & Integration of Specialized Knowledge & Knowledge Management and Teamwork & [S57], [S61] \\ \hline
\normalsize
\end{tabular}
\end{table}

\section{Discussion} \label{sec:discussao}

The results indicate that communication and teamwork are the most frequently mentioned soft skills in the analyzed publications, consolidating themselves as core competencies in development teams that adopt agile practices. However, skills such as proactivity, organization, and responsibility appear less frequently, suggesting that the literature still tends to focus oclassifying communication as a soft skill that is both stable, given its continuous presence over time, and emerging, umber of studies and in the identification of soft skills. The temporal analysis shows that communication has remained the most recurrent soft skill since 2003, with a marked increase from 2020 onwards, possibly due to the growth of agile practices and remote work.

In the definition of activities and processes, the findings highlight communication as an essential skill for facilitating interactions with clients, users, and stakeholders, aiming to understand needs and expectations. The lack of communication may lead to project failure \cite{antlova_agile_2014, hidellaarachchi_influence_2023}. Ceremonies such as Sprint Review and Daily Scrum emphasize this skill \cite{baumgart2015personality}. In addition to its high frequency, communication consistently appears throughout the entire analyzed period, never disappearing from the literature. Its sustained presence reflects the very nature of agile methodologies, which are grounded in frequent interactions, continuous feedback, and constant alignment among team members, clients, and other stakeholders. In this sense, communication underpins and amplifies other competencies, such as teamwork, problem solving, and adaptability. Consequently, the findings support the classification of communication as a soft skill that is simultaneously stable due to its continuous presence over time and emerging given its significant growth in recent years. These characteristics reinforce its structural relevance in agile contexts and indicateclosely linked to the ability to respond to changing requirements, replan activities, and incorporate continuous client and stakeholder feedback teams. Teamwork is a pillar of agile methodologies and is fundamental for aligning expectations, sharing knowledge, and supporting peers \cite{baumgart2015personality, exter_comparing_2018, lima2019, lucianosilva2023}. Practices such as pair programming rely heavily on collaboration \cite{kovitz_hidden_2003, fagerholm_performance_2015, matturro_soft_2015}. Leadership in agile teams is not restricted to formal positions and may be shared \cite{fagerholm_how_2014, miranda2021}. The Scrum Master, in particular, requires leadership skills to facilitate communication and guide the team.

Adaptability is central to agile methodologies, enabling teams to deal with frequent changes. The temporal analysis of the studies reveals a consistent growth of this skill over the years, with a significant intensification from 2020 onwards. This increase coincides with the COVID-19 pandemic period, during which teams had to deal with abrupt transformations in work processes, communication practices, and task organization, reinforcing the need for professionals capable of rapidly adjusting to new contexts, technologies, and organizational demands. In agile environments, adaptability is directly associated with the ability to respond to changing requirements, replan activities, and incorporate continuous feedback from clients and stakeholders. Unlike traditional models, agile methodologies emphasize flexibility and constant iteration, requiring team members to deal with uncertainty in a proactive and collaborative manner. Thus, adaptability is not limited to reacting to change but involves an active stance toward learning, experimentation, and openness to redefining practices and roles throughout the development lifecycle. The data indicate that adaptability is more frequently emphasized in roles directly involved in execution and coordination, such as Developers, Scrum Masters, and QA/Test Analysts, who regularly face technical, organizational, and scope-related changes. This recurrence suggests that adaptability functions as a transversal competence, supporting other soft skills such as communication, problem solving, and teamwork \cite{fagerholm_how_2014, gren_non-technical_2018, machado2021}. Professionals must be flexible and open to changes in plans, technologies, and processes. Proactivity is expected in self-organizing teams, allowing members to identify and execute tasks without constant supervision \cite{matturro_soft_2015, mangiza_requisite_2020, florea_roles_2023, lucianosilva2023}. Motivation is associated with performance and commitment to project goals \cite{ciric_importance_2020}. Organization and responsibility are central to planning, scope definition, and task prioritization, as exemplified by the Product Owner role in backlog management \cite{baumgart2015personality, matturro_soft_2015, assyne_essential_2022}.

Regarding methodologies, Scrum appears in 21\% of the publications, followed by combinations of Scrum and XP (15\%). Other methodologies, such as SAFe, appear less frequently. More than half of the studies (53.4\%) do not specify the methodology used, which limits result comparability. Scrum tends to reinforce competencies related to communication and teamwork, whereas XP emphasizes problem solving and feedback, associated with practices such as pair programming and test-driven development.

The role-based analysis shows that communication is attributed to virtually all profiles, with higher incidence among Scrum Masters and Technical Leaders due to their responsibilities in conflict mediation, strategic alignment, and ceremony facilitation. Most studies (64 publications, 65.9\%) do not specify seniority levels. Among those that do, references to all levels (20, 20.6\%) and senior profiles (5, 5.15\%) predominate. Mentions of junior, mid-level, or intern roles are rare, indicating a gap in the literature regarding the differentiation of soft skills across career stages.

Overall, a recent strengthening of socio-emotional soft skills can be observed in the literature, with emphasis on empathy, emotional intelligence, active listening, and conflict management, particularly in the most recent years of the analyzed period. This shift suggests a growing research focus on human-centered aspects of agile work, possibly associated with the consolidation of remote and hybrid environments, increasing team diversity, and heightened concerns about well-being and collaboration in distributed settings. In this context, agile software development is increasingly understood as a complex socio-technical activity in which human and emotional factors play a central role. At the same time, the results indicate the existence of a core set of soft skills broadly distributed across roles alongside a group of competencies more closely associated with specific responsibilities, including leadership, organization, and goal setting. This pattern highlights that, while agile teams rely on shared interpersonal competencies, their effectiveness also depends on strengthening role-aligned soft skills. Consequently, the findings reinforce the importance of differentiated approaches to soft skill development, avoiding both excessive generalization and rigid specialization, and establish a relevant bridge for future investigations into mental health, hybrid teams, and the impact of emerging technologies, such as Artificial Intelligence, on the mediation of interactions and work dynamics.

\section{Guidelines for the Agile Community}

This section presents guidelines derived from this SMS, aimed at supporting the agile community across education, industry, and research. Beyond merely synthesizing the identified soft skills, the proposed guidelines translate empirical evidence consolidated over 25 years of literature into practical directions that can inform curriculum design, improve organizational practices, and guide future scientific investigations.

\subsection{Education}

Based on patterns observed over 25 years of literature, the results point to clear directions for professional education, industry practices, and the advancement of scientific research in Computer Science. A core set of interpersonal skills (Communication, Teamwork, Adaptability, and Leadership) shows higher frequency within development teams, regardless of the role performed in agile environments. This pattern suggests that these skills should not be treated as complementary or optional, but rather as foundational competencies in the education of computing professionals.

Accordingly, the findings underscore the need to rethink curricula and pedagogical strategies by intentionally integrating soft skill development throughout undergraduate and graduate programs. Approaches such as Project-Based Learning (PBL), collaborative and multidisciplinary activities, agile environment simulations, and formative assessment are particularly effective in fostering these competencies. Moreover, given the rapid evolution of Information Technology, the results emphasize the importance of practical courses and continuous training focused on socio-emotional skills to support professionals’ adaptation to technological and organizational change. Less frequently discussed skills remain essential in professional practice, revealing educational gaps that may be addressed through learning experiences grounded in real-world contexts.

\subsection{Industry}

From an industry perspective, the results provide concrete support for improving organizational processes, particularly in team composition, recruitment and selection, and agile development management. The recurrence of specific soft skills associated with particular roles (e.g., Communication and Leadership for Scrum Masters and Technical Leaders, or Adaptability and Problem Solving for Developers and QA professionals) indicates that these competencies can be used as more objective criteria in interviews and performance evaluations.

Moreover, the results suggest that organizations may benefit from aligning agile practices with human development strategies by promoting training and activities that strengthen behavioral skills critical to team success. The identification of non-technical skills widely distributed across roles also reinforces the importance of collaborative and psychologically safe environments in which communication, feedback, and continuous learning are systematically encouraged.

\subsection{Research}

In the research context, the findings reveal relevant gaps and opportunities for future investigation. Despite the broad identification of skills across the literature, many studies do not explicitly differentiate skills by role, seniority level, or agile methodology, limiting more refined analyses. This scenario suggests the need for empirical studies that deepen the understanding of role-specific soft skills and their evolution across professional careers. Furthermore, it highlights the importance of investigating how soft skills are effectively developed, assessed, and integrated into agile practices, moving beyond simple identification. Future research may explore causal relationships among soft skills, performance indicators, software quality, and team well-being, and may propose models, frameworks, or tools to support the systematic incorporation of these competencies into software development.

\section{Threats to Validity} \label{sec:ameacas}

During the conduct of this SMS, several threats to validity were identified and analyzed using the frameworks proposed by Lago et al. \cite{lago2024threats} and Ampatzoglou et al. \cite{ampatzoglou2020guidelines}. To mitigate these threats, methodological strategies were adopted throughout the planning, execution, data extraction, and analysis stages. The threats are organized into three main categories: study selection validity, data validity, and research validity.

Study selection validity concerns threats arising during the identification, search, and filtering of primary studies. A key threat relates to publication identification, particularly the construction of the search string, the selection of digital libraries, and the definition of the temporal scope. An incomplete or overly restrictive search string may lead to the exclusion of relevant studies. This risk was mitigated through a pilot test of the search string and the prioritization of well-established digital libraries in Computer Science and Software Engineering.

Another threat is associated with study accessibility and language. The exclusion of articles not written in English, Portuguese, or Spanish, as well as studies without full-text access, may introduce availability bias and reduce coverage. Although this limits study diversity, it was adopted to ensure consistent analysis and reliable data verification. Inclusion and exclusion criteria may also introduce bias when subjectively applied; therefore, they were clearly defined and discussed among the authors whenever uncertainties arose.

Data validity includes threats related to data extraction and analysis. Data extraction bias may occur due to the researcher's subjectivity when interpreting study results. To mitigate this risk, a predefined extraction form was used, and the process was reviewed by multiple authors. Publication bias also represents a relevant threat, as studies with positive results are more likely to be published. Although this bias cannot be fully eliminated, searching multiple databases helps reduce its impact. Additionally, limited sample size and study heterogeneity may weaken conclusions; thus, results were analyzed descriptively and qualitatively.

Research validity encompasses threats related to the overall study design. Repeatability may be compromised if the review protocol lacks sufficient detail. To address this issue, all stages of the process were documented and made publicly available \cite{repositorio}. Another threat concerns the coverage of the research questions, which may lead to partial results if poorly defined. This risk was mitigated by grounding the research questions in identified literature gaps. Finally, the generalizability of the findings may be limited, as the results reflect specific software industry and agile methodology contexts. Therefore, their application to other domains should be approached with caution.

\section{Conclusion} \label{sec:conclusao}

This study highlights the growing relevance of intrapersonal and interpersonal competencies for computing professionals in contemporary software development contexts. By analyzing 97 primary studies, this SMS identified 33 distinct soft skills, revealing a consistent core set widely recognized as fundamental in agile environments.

The results show that, beyond technical expertise, the effectiveness of agile teams is strongly influenced by socio-emotional capabilities that support collaboration, adaptability to change, and shared responsibility. While these essential soft skills are broadly distributed across roles, others (such as organization, leadership, and goal setting) are more closely associated with specific responsibilities, reinforcing the need for role-aware approaches to skill development.

This study also exposes important research gaps. Limited attention has been given to emerging and specialized roles, such as DevOps engineers, designers, and data scientists, indicating that the diversity of modern software teams is not yet fully represented in the literature. Future research should examine how soft skills are developed and sustained over time, including the role of structured interventions such as training, mentoring, and organizational support. In addition, the increasing integration of Artificial Intelligence (AI) into software development raises questions about how soft skills are reshaped in AI-mediated environments.

\section*{Artifacts Availability}

All artifacts from this study are available at \url{https://zenodo.org/records/19899249}


\section*{Acknowledgments}
The authors gratefully acknowledge the financial support provided for this research. Márcio Ribeiro was partially supported by INES.IA (National Institute of Science and Technology for Software Engineering Based on and for Artificial Intelligence) (www.ines.org.br), under CNPq grant 408817/2024-0, as well as by CNPq grants 312195/ 2021-4, 404825/2023-0, and 443393/2023-0. Ivan Machado acknowledges partial support from the CAPES – Finance Code 001; Fundação de Amparo à Pesquisa do Estado da Bahia (FAPESB), grant PIE0002/2022; and CNPq, grants 315840/2023-4 and 403361/2023-0. Carla Bezerra is supported by 403304/2025-3.

\section*{Use of Artificial Intelligence Tools}
The authors acknowledge the use of OpenAI’s GPT-5.3 language model for grammatical revision and translation of this manuscript. The authors retain full responsibility for the content and intellectual contributions presented.
\bibliographystyle{ACM-Reference-Format}
\bibliography{sample-base}

\end{document}